\documentclass[%
 aip,
pop,
 amsmath,amssymb,
preprint,%
]{revtex4-1}

\usepackage{graphicx}
\usepackage{dcolumn}
\usepackage{bm}

\usepackage[utf8]{inputenc}
\usepackage[T1]{fontenc}
\usepackage{mathptmx}
\usepackage{etoolbox}

\makeatletter
\def\@email#1#2{%
 \endgroup
 \patchcmd{\titleblock@produce}
  {\frontmatter@RRAPformat}
  {\frontmatter@RRAPformat{\produce@RRAP{*#1\href{mailto:#2}{#2}}}\frontmatter@RRAPformat}
  {}{}
}%
\usepackage{graphicx}

\usepackage[utf8]{inputenc}
\usepackage[T1]{fontenc}
\usepackage{amsmath}
\usepackage{etoolbox}

\patchcmd{\subequations}{\alph{equation}}{\textit{\alph{equation}}}{}{}

\usepackage{cleveref}
\usepackage{bm}
\usepackage{mathtools}

\usepackage{physics}

\newcommand{\R}{\bm R}

\newcommand{\N}{\mathcal{N}}

\renewcommand{\|}{\parallel}

\newcommand{\projpj}[2]{ \norm{#2}^{#1}}

\renewcommand{\poissonbracket}[2]{  \left[#1, #2 \right]}

\renewcommand{\eqref}[1]{Eq. (\ref{#1})}

\newcommand{\secref}[1]{Sec. \ref{#1}}
\newcommand{\figref}[1]{Fig. \ref{#1}}

\newcommand{\upare}{U_{\parallel e}}
\newcommand{\del}{\partial}
\newcommand{\uwall}{U_{\|}(\text{wall}_{\pm})}
\newcommand{\duwall}{\del_z \uwall}

\makeatother
\begin{document}

\preprint{AIP/123-QED}

\title[Sheath boundary conditions in a linear plasma device]{Derivation and application of sheath boundary conditions for drift-kinetic simulations in a linear plasma device based on a gyromoment approach}
\author{J. E. Mencke}
 \email{jacob.mencke@epfl.ch}

\author{T. Stucker}%
\author{P. Ricci}%
\affiliation{ 
Ecole Polytechnique F\'ed\'erale de Lausanne (EPFL), Swiss Plasma Center, CH-1015 Lausanne, Switzerland
}%

\date{\today}

\begin{abstract}
    Boundary conditions for a drift-kinetic model at the collisional presheath entrance with perpendicular incidence of the magnetic field to the wall are derived and numerically implemented. The drift-kinetic model for the plasma is based on the expansion of the ion distribution function on a Hermite-Laguerre basis, and the evolution of the resulting gyromoments. A linear-plasma-device geometry is considered. Comparison with simpler simulations with previously used ad hoc boundary conditions is presented. For the new set of boundary conditions, a significant increase of the plasma outflow to the wall is observed, leading to a significantly smaller plasma density in the whole volume of the device.
\end{abstract}

\maketitle

\section{Introduction}\label{sec:introduction}
Drift-reduced fluid \cite{braginskii1965transport,Zeiler1997}, drift-kinetic (DK) \cite{catto2004drift,Jorge2017drift}, and gyrokinetic (GK) \cite{littlejohn1983variational,Brizard2007,hahm2009fully} models rely on a separation of scales to efficiently describe magnetized plasmas. The gyromotion is averaged out and length scales much larger than the Debye length are considered, such that the plasma can be considered quasi-neutral. However, these approximations break down in the proximity of solid walls. At a few Debye lengths away from the wall, the quasi-neutrality assumption is not satisfied. This region is defined as the Debye sheath \cite{Chodura1982,Riemann_1991}. If the incidence of the magnetic field with the solid wall is oblique, at a distance of the order of the ion gyro-orbit, typically larger than the Debye length in fusion plasmas, the ion trajectories intersect the wall defining the entrance of the magnetic presheath. In the magnetic presheath, as well as in the Debye sheath, due to the large electric field, the gyro-orbit is not well defined, and the magnetic moment is not conserved \cite{franklin2003plasma}. Finally, even if the the plasma is dense and cold in the vicinity of the wall, such as in the case of basic plasma experiments or in detached conditions in fusion devices, collisions become negligible at a few mean-free-path lengths away from the wall, defining the collisional presheath entrance. To avoid compromising the drift-reduced fluid, DK, or GK ordering assumptions to study the dynamics of the bulk plasma, boundary conditions (BCs) need to be imposed at the entrance of the collisional presheath, magnetic presheath, or Debye sheath where these assumptions are first violated. These BCs should properly capture the dynamics inside the sheath providing its physics to the bulk plasma model \cite{Chodura1982,riemann1994theory,stangeby2000tutorial,loizu2011existence,Loizu2012}.



The interplay between the plasma sheath and bulk plasma is highly non-trivial and several approaches have been considered to provide proper BCs. For fluid models, the Bohm-Chodura BC is usually imposed, requiring that the fluid outflows at the sound speed at the collisional presheath entrance \cite{Stangeby1995,Loizu2012,Giacomin2022}. For GK simulations, logical sheath BCs are commonly used. These state that the electrons with insufficient velocity to overcome the sheath potential barrier are reflected, while all ions are absorbed \cite{Parker1992,ku2018fast,bottino2025particle}. Despite the success of these types of BCs, more detailed treatments suggest that the buildup of large electric fields at the sheath entrance can drastically alter the BCs at a shallow-angle incidence of the magnetic field with the wall \cite{Geraldini2017,geraldini2019dependence}.

In this work, the bulk plasma is assumed to satisfy a DK ordering in the electrostatic limit. The turbulent fluctuations are allowed to be large in amplitude and develop on large scales with respect to the ion Larmor radius. The ion distribution function is projected on a Hermite-Laguerre basis and the DK moment hierarchy from \cite{mencke_full_delta} is utilized. By imposing a critical balance ordering, the electrons are described using a drift-reduced Braginskii model \cite{mencke_full_delta,braginskii1965transport,FelixParraBrag}. The magnetic geometry we consider is that of a linear plasma device, and we focus on the BCs at the end of the plasma column, thus assuming that the incidence of the magnetic field to the wall is perpendicular. This significantly simplifies the treatment of the boundary conditions as no magnetic presheath is present. Analytical BCs for all fields are consistently derived assuming that the plasma is in a collision-dominated regime in the proximity of the wall, and therefore that a collisional presheath is present. The BCs are thus imposed at the collisional presheath entrance. In tokamak experiments, these conditions can be achieved for a diverted plasma in detached conditions, for example due to heavy impurity seeding \cite{Fevrier_2020}. For the case of basic plasma-physics experiments, such as the LAPD linear plasma device, the entire bulk plasma is highly collisional \cite{Gekelman1991,mencke_full_delta}. The approach we use to derive the BCs, is similar to previous work where boundary conditions are derived for a two-fluid model \cite{Loizu2012,loizu2011existence}. Despite the simple geometry, our treatment provides insight into how sheath BCs affect the behavior of global turbulent plasma simulations. 

This paper is structured as follows. In section \secref{sec:oredering_field_equations} we summarize the ordering and governing equations used in Ref. \cite{mencke_full_delta} and in this work as well. In \secref{sec:bc}, we derive a new set of BCs at the collisional presheath entrance. The numerical implementation of the model and BCs is described in \secref{sec:BC_impl}. We present the simulation results in \secref{sec:sim_res} and compare with the DK simulation in \cite{mencke_full_delta} where standard Bohm BCs for the parallel flow and homogeneous Neumann BCs for all other fields are imposed. Finally, \secref{sec:concl} concludes the paper.
 
\section{Ordering, governing equations, and numerical scheme}\label{sec:oredering_field_equations}
We consider a fully ionized plasma in a linear plasma device, and adopt the electrostatic approximation. We order the perpendicular gradient lengths, $k_{\perp }\sim \left|\nabla_{\perp} \ln F_{a}\right|\sim \left|\nabla_{\perp} \ln \phi \right|\sim \epsilon_{\perp}/\rho_{s}$, with $\epsilon_\perp<1$, being $\rho_s=c_s/\Omega_i$ the ion sound Larmor radius with $c_s=\sqrt{T_e/m_i}$ the sound speed, $T_e$ the electron temperature, and $\Omega_i=q_i B/m_i$ the ion gyro frequency with $q_i$ the charge of the ions. Furthermore, we order the parallel length scale, $k_{\|}\sim \epsilon k_{\perp }$, and frequency scale, $\omega\sim \Omega_i \epsilon\epsilon_{\perp}$, with $\epsilon<1$. This ordering is appropriate for conditions in the scrape-off layer of fusion plasmas, where the density and temperature fluctuate on large scales with order one amplitude \cite{Zweben_2015,Nespoli_2017}.


A plasma model that satisfies this ordering is the electrostatic DK system derived in \cite{mencke_full_delta}, summarized here. For the ions, the gyromoments, $\N^{pj}_{i}$, are introduced as

\begin{subequations}
\begin{align}
    \N^{pj}_i\left(\R,t\right) &=2\pi\int_{-\infty}^{\infty}dv_{\|}\int_0^\infty d\mu \frac{B}{m_i}F_i\left(\R, v_{\|},\mu,t\right) \frac{H_{p}\left(s_{\| i}\right)L_{j}\left(x_i\right)}{\sqrt{2^{p}p!}}  ,\\
\intertext{where we define}
    s_{\| i}&=\frac{v_{\|}-U_{\| i}\left(\R,t\right)}{v_{Th i}},\quad x_i = \frac{\mu B}{T_{i0}},\quad v_{Th i}=\sqrt{\frac{2T_{i0}}{m_i}},
\intertext{and where $H_p$ is the physicist's Hermite polynomial of order $p$, $L_j$ is the Laguerre polynomial of order $j$, with $v_{Thi}=\sqrt{2T_{i0}/m_i}$, and $T_{i0}$ is a reference ion temperature. With these definitions, the ion distribution function is given by}
    F_i\left(\R,v_{\|}, \mu.t\right)&=\sum_{p,j=0}^{\infty}\N^{pj}_{i}\left(\R,t\right)\frac{H_{p}\left(s_{\| i}\right)L_{j}\left(x_i\right)}{\sqrt{2^{p}p!}} F_{Mi}\left(s_{\| i},x_i\right),\label{eq:DK_Npj_exp}
\intertext{with}
    F_{Mi}&=\frac{e^{-s_{\|i}^2-x_i}}{\left(\sqrt{\pi}v_{Thi}\right)^3}.
\end{align}
\end{subequations}
The evolution equations for the gyromoments are
\begin{subequations}\label{eq:Mother_eq}
\begin{align}
    \partial_t \N^{pj}_{i}&+ \nabla_{\|}\projpj{pj}{v_{\|}}_{i}+\sqrt{p}\projpj{p-1 j}{s_{\| i}}_{i} \nabla_{\|}U_{\|}+\frac{1}{B}\poissonbracket{\phi}{\N^{pj}_{i}}\nonumber\\
        &- \frac{\sqrt{2p}\N^{p-1 j}_{i}}{N_{e}}\frac{1}{m_i v_{Thi}}\nabla_{\|} P_{\| i }=  C_{ii}^{pj} + S^{pj}_{i},\label{eq:moment_hierachy_DK}
\intertext{where the projector of a phase-space function, $\chi$, is defined as}
        \projpj{pj}{\chi}_i&= 2\pi \int_{-\infty}^{\infty} dv_{\|}\int_{0}^{\infty} d\mu \frac{H_{p}\left(s_{\| i}\right)L_{j}\left(x_i\right)}{\sqrt{2^{p}p!}}\frac{B}{m_i} F_{i}\chi,
\intertext{and the short-hand notation is used in the following,}
    \projpj{}{\chi}_i&=\projpj{00}{\chi}_i.
\intertext{We thus have}    
        \projpj{pj}{s_{\|}}_i& =  \sqrt{p+1}\N^{p+1 j}_{i}+\sqrt{p}\N^{p-1 j}_{i},
\intertext{and}
        \projpj{pj}{v_{\|}}_i&= \frac{v_{Th i}}{\sqrt{2}}\projpj{pj}{s_{\|}}_i+U_{\|} \N^{pj}_i.
\intertext{On the other hand, assuming $\omega\sim \nu_{ii}$ with $\nu_{ii}$ being the ion-ion collision frequency, the electron-electron collision frequency, $\nu_{ee}\sim \sqrt{m_i/m_e}\nu_{ii}\gg \omega$, is large thus we consider a fluid model for the electrons. By considering quasi-neutrality, the evolution of the electron density, $N_e$, is given by}
        \partial_t  N_{e}&+\frac{1}{B}\poissonbracket{\phi}{ N_{e}}+\nabla_{\|}\left(U_{\|}N_{e}\right)=S_N\label{eq:dt_Ne}.
\intertext{The evolution of the electron temperature is}
         \partial_t  T_{e}&+\frac{1}{B}\poissonbracket{\phi}{ T_{e}}+U_{\| }\nabla_{\|}T_{e}+\frac{2}{3}T_{e}\nabla_{\|} U_{\| }+\frac{2}{3N_{e}}\nabla_{\|}q_{\| e}-0.71\frac{2T_e}{3N_e}\nabla_{\|}J_{\|}=S_{Te},\label{eq:dt_Te}
\intertext{with}
    q_{\| e}&=-\frac{3.16N_{e}T_{e}}{m_e\nu_{ee}}\nabla_{\|}T_{e},
\intertext{and the equation for the parallel velocity, $ U_{\|}$, defined by}
    m_{i}U_{\|}&\equiv m_iU_{\| i}+m_e U_{\| e}
\intertext{is given by}
    m_iN_{e}&\left(\partial_{t} U_{\| }+ U_{\| }\nabla_{\|} U_{\| }+\frac{1}{B}\poissonbracket{\phi}{U_{\| }}\right) +\nabla_{\|} P_{\| i}+\nabla_{\|} P_{\| e} =0,\label{eq:Upar}
\intertext{with the ion parallel velocity defined as}
    U_{\| i}&=\projpj{}{v_{\|}}_i/N_{i},\label{eq:Upara_DK}
\intertext{the ion parallel pressure}
    P_{\| i}&=m_i\projpj{}{\left(v_{\|}-U_{\| }\right)^2}_i=T_{i0}\left(\sqrt{2}\N^{20}_{i}+N_{e}\right), \label{eq:Pparallel}
\intertext{and the ion perpendicular pressure}
    P_{\perp i}&=\projpj{}{\mu B}_i=T_{i0}\left(N_{e}-\N^{01}_{i}\right). \label{eq:Pperp}
\intertext{For closing the system, we use a vorticity equation, that is}
    \partial_t\Omega&=\nabla_{\|} J_{\|}-\nabla_{\|}\left(\Omega U_{\| }\right)-\frac{1}{B}\poissonbracket{\phi}{\Omega}-\frac{1}{2m_i\Omega_i^2}\nabla_{\|}\left(P_{\perp } \nabla_{\perp}^2 U_{\| }\right)\nonumber\\
    &-\frac{1}{B}\poissonbracket{\nabla_{\perp}\phi\cdot}{\boldsymbol{\omega} }-\nabla_{\|}\left(\boldsymbol{\omega}\cdot\nabla_{\perp}U_{\| }\right)+\frac{1}{B\Omega_i}\nabla_{\perp}S_N\cdot\nabla_{\perp} \phi\nonumber\\
    &+\frac{1}{2m_i\Omega_i^2}\poissonbracket{\nabla_\perp^2\phi}{P_{\perp }}-\frac{1}{2m_i\Omega_i^2}\nabla_{\|}\left(\nabla_{\perp}^2 Q_{\perp }\right),\label{eq:dt_Omega}\\
    \intertext{with}
    \boldsymbol{\omega}&=\frac{1}{B\Omega_i}N_{e}\nabla_{\perp}\phi+\frac{1}{m_i\Omega_i^2}\nabla_{\perp}P_{\perp },\quad Q_{\perp i}=\projpj{}{v_{\|}\mu B}
    \intertext{and}
    \Omega&=\nabla_{\perp}\cdot\boldsymbol{\omega}\label{eq:Omega_impl},
\intertext{as well as an equation for the parallel current given by}
     \partial_t J_{\|}&+\frac{1}{B}\poissonbracket{\phi}{J_{\|}}+\nabla_{\|}\left(J_{\|} U_{\|}\right)+J_{\|}\nabla_{\|} U_{\|}+\frac{e}{m_e}N_{e}\nabla_{\|} \phi+\frac{1}{m_i}\nabla_{\|} P_{\| i}\nonumber\\
    &-\frac{1}{m_e}\nabla_{\|}\left(N_{e}T_{e}\right)=\frac{S_N}{N_{e}}J_{\|}-\nu_{\|} J_{\|}+0.71N_{e}\nabla_{\|}\left(\frac{1}{m_e}\nabla_{\|}T_{e}\right)+S_J.\label{eq:dt_J}
\intertext{Furthermore, we note that $\nu_{\|}=4\sqrt{2\pi} e^4 N_{e}\sqrt{m_e} \ln \Lambda /[3   m_i T_{e}^{3/2}1.96  \left(4\pi \epsilon_0\right)^2]$ is the Spitzer resistivity and $\nu_{ee}=m_i/m_e 1.38\nu_{\|}$ is the electron-electron collision frequency with $ \ln\Lambda$ being the Coulomb logarithm. A Dougherty collision operator is used in \eqref{eq:moment_hierachy_DK}, with projection given by \cite{frei2023fullf,mencke2025extended,mencke_full_delta}}
    C_{ii}^{pj}&=\nu_{ii}\left[-\left(p+2j\right)\N^{pj}_{i}+\left(T_{i}-1\right)\vphantom{\left(\sqrt{p\left(p-1\right)}\N^{p-2 j}_{i}-2j \N^{pj-1}_{i}\right)}\left(\sqrt{p\left(p-1\right)}\N^{p-2 j}_{i}-2j \N^{pj-1}_{i}\right)\right],\label{eq:C_iDK}
\intertext{being $\nu_{ii}=\nu_{ee}\sqrt{m_e/m_i}\left(T_{e}/T_{i}\right)^{-3/2}$. We note that the choice of the Dougherty collision operator is motivated by its particles and energy conservation properties and its a simple projection on the Hermite-Laguerre basis. We choose the source terms to have the following form}
    S_{i}^{pj}&=A_{N} \delta^{pj}_{00}+A_{E} \left(\frac{\delta_{20}^{pj}}{\sqrt{2}}-\delta^{pj}_{01}\right)N_{eDK}F_{Mi},\\
    A_{N}&=S_N= \frac{\mathcal{A}_{N0}}{2}    \left[ 1 - \tanh\left( \frac{r -r_s}{L_s}\right)  \right] + \mathcal{A}_{N \infty},
\end{align}
\end{subequations}
The energy and temperature sources, $A_E$ and $S_{Te}$, have an equivalent definition as $A_N$ with $A_{Te0}$, $A_{Te\infty}$, $A_{E0}$, and $A_{E\infty}$. In order to advance the plasma dynamics, we evolve $N_e$, $T_e$, $\N^{pj}$, $U_{\|}$, $\Omega$ and $J_{\|}$ in time and invert \eqref{eq:Omega_impl} to obtain $\phi$.

In the following, the time, $t$, is normalized to $ R / c_{s0}$, being $c_{s0} = \sqrt{T_{e0} / m_i}$ the ion sound speed at the constant reference electron temperature, $T_{e0}$, while $R$ is the characteristic length parallel to the magnetic field, comparable to the size of the experiment, defined here as the perpendicular radius of the plasma chamber. The electrostatic potential, $\phi$, is normalized to $ T_{e0} / e$, the parallel spatial scales to $R$, and the perpendicular ones to $\rho_{s0} = c_{s0} / \Omega_i$. Both the gyromoments, $\N^{pj}_i$,  and the electron density $N_e$, are normalized to the constant reference density $N_{0}$, the parallel fluid velocities, $U_{\parallel }$ to $c_{s0}$, the electron and ion temperatures, $T_e$, $T_{\| i}$, and $T_{\perp i}$, to $T_{e0}$, $T_{i0}$, and $T_{i0}$, respectively. Finally, we normalize the parallel current density, $J_{\|}$, to $N_{0}c_{s0}$.

\section{Derivation of boundary conditions}\label{sec:bc}

In order to derive appropriate sheath BCs for the model in \eqref{eq:Mother_eq}, we assume the presence of a cold and dense plasma in the vicinity of the wall. Furthermore, we assume that the ion-mean-free-path length, $\lambda_{\| i}=\sqrt{2T_i/m_i}/\nu_{ii}$, is much larger than the Debye length but much smaller than the parallel gradient length scale, $\lambda_{D}\ll \lambda_{\| i}\ll R$. Thus, we impose our BCs at the collisional presheath entrance, where the distance to the wall is comparable to the ion-mean-free path. In agreement with \cite{Loizu2012}, to derive the BCs, we assume that:

\begin{enumerate}
    \item time scales within the sheath are much faster than in the bulk plasma, implying that the sheath is considered being in steady-state $\partial_t \approx 0$;
    \item parallel gradients dominate over perpendicular gradients and source terms, due to the strong electric fields in the parallel direction close to the walls; 
    \item the $\nabla_{\|} q_{\| e}$ term in \eqref{eq:dt_Te} is negligible;
    \item electrons are distributed according to a cut-off Maxwellian, the cut-off representing the sheath potential reflecting all electrons with parallel velocity below a certain cut-off velocity \cite{Loizu2012}, such that the electron fluid velocity is expressed as
    \begin{equation} \label{eq:upare}
        \upare = \sqrt{\frac{T_{e}}{m_i}}\exp \left(\Lambda - \frac{\phi}{T_e} \right);
    \end{equation}
    \item a high collisionality regime for the ions ($\nu_{ii} \gg \omega$);
    \item negligible electron inertia, $m_e/m_i\sim 0$.
\end{enumerate}



Since the plasma is collisional at the presheath entrance,  the high-collisionality limit to close our moment hierarchy in \eqref{eq:moment_hierachy_DK} is assumed by imposing $C_{ii}^{pj} = 0$ for all $p$ and $j$. This implies that for all moments with $\left(p,j\right)\neq \left(0,0\right)$, the gyromoments are expressed by the recursive relation
\begin{equation} \label{eq:Npjrecu}
    \N^{pj}_i = \frac{T_i - 1}{p + 2j} \left[ \sqrt{p(p-1)} \N^{p-2j}_i - 2j \N^{pj-1}_i \right].
\end{equation}
We note that the relation in \eqref{eq:Npjrecu} is valid for a Maxwellian distribution function. Indeed, projecting a local Maxwellian,
\begin{subequations}
    \begin{align}
        F_{Mli}&=N_{i}\left(\frac{m_i}{2\pi T_{i}}\right)^{3/2}\exp\left(-\frac{\mu B}{T_{i}}-\frac{m_i\left(v_{\|}-U_{\| i}\right)^2}{2T_i}\right),
\intertext{on the Hermite-Laguerre basis in \eqref{eq:DK_Npj_exp} yields}
        \N^{pj}_{Ml}&=2\pi \int_{-\infty}^{\infty}dv_{\|}\int_{0}^{\infty}d\mu \frac{B}{m_i} F_{Mli}\frac{H_{p}\left(s_{\| i}\right)L_{j}\left(x_i\right)}{\sqrt{2^{p}p!}} \nonumber\\
\intertext{and by devoloping the integrals, one obtains}
        \N^{pj}_{Ml}&=\begin{cases}
        N_i\frac{\left(-1\right)^{j}\sqrt{p!}}{\sqrt{2^{p}}\left(p/2\right)!}\left(\frac{T_{ i}}{T_{i0}}-1\right)^{p/2+j}& p\:\text{even,}\\
        0 & p\:\text{odd,}
    \end{cases}\label{eq:MaxNpj}
    \end{align}
\end{subequations}
as a consequence of the particle and energy conserving properties of the Dougherty collision operator.

Since $\N^{10}_i = 0$ \cite{Frei2020,frei2023fullf,mencke2025extended,mencke_full_delta}, we find from \eqref{eq:Npjrecu} that $\N^{pj}_i = 0$ for all odd $p$'s. We thus deduce that homogeneous Dirichlet BCs should be applied to the gyro-moments with odd $p$. Furthermore, we notice that setting $\left(p,j\right)=\left(2,0\right)$ in \eqref{eq:Npjrecu} and using \eqref{eq:Pparallel}, we obtain $P_{\| i}=P_{i}$. Similarly, setting $\left(p,j\right)=\left(0,1\right)$ in \eqref{eq:Npjrecu} and using \eqref{eq:Pperp}, we get $P_{\perp i}=P_{i}$. Then, \eqref{eq:Npjrecu} recursively relates all moments with $p>2$ and/or $j>1$  to the moments with smaller $p$ and $j$. Thus, once a set of BCs for the moments with $\left(p,j\right)=\left(0,0\right)$, $\left(2,0\right)$, and $\left(0,1\right)$ is derived, boundary conditions for all other moments can be obtained using \eqref{eq:Npjrecu}.


To derive the BCs for the moments $\left(p,j\right)=\left(0,0\right)$, $\left(2,0\right)$, and $\left(0,1\right)$, we first notice that the ion temperature can be written as a linear combination of the gyromoments as
\begin{equation}
    T_i = \frac{2T_{\perp i} + T_{\parallel i}}{3} = T_{i0}\frac{\sqrt{2}\N^{20}_i + 3N_e - 2\N^{01}_i}{3N_e}.\label{eq:T_i}
\end{equation}

Combining  \eqref{eq:moment_hierachy_DK} with $\left(p,j\right)=\left(0,0\right)$, $\left(2,0\right)$, and $\left(0,1\right)$, and using \eqref{eq:T_i}, we get
\begin{align}
    \partial_t \left(N_e T_i\right)& +\nabla_{\|}\left(N_eT_iU_{\|}\right)+\frac{2}{3}T_{i}N_e\nabla_{\| }U_{\|}+\frac{1}{\rho_*}\poissonbracket{\phi}{N_e T_i}\nonumber\\
    &=\frac{T_{i0}}{3}\left(\sqrt{2}S^{20}_i+3S^{00}_{i}-2S^{01}_{i}\right),\label{eq:dt_Pi}
\end{align}
where we have used that, at high collisionality, $P_{\| i}=N_e T_i$.

We rewrite the electron velocity in terms of the current density as follows
\begin{subequations}
\begin{align}
    U_{\| e}&=U_{\| i}-\frac{J_{\|}}{N_e}\sim U_{\|}-\frac{J_{\|}}{N_e},
\intertext{and since $J_{\|}\sim \epsilon_{\perp}^2N_e c_{s}$ \cite{mencke_full_delta}, we assume here that}
    U_{\| e}&\sim U_{\|} \label{eq:Upare}.
\end{align}
\end{subequations}
Therefore, we replace $U_{\|}$ with $U_{\| e}$ in \eqref{eq:dt_Te} in the reminder of this chapter.



By only keeping first order parallel derivatives, we use Eqs. (\ref{eq:Upar}), (\ref{eq:dt_Ne}), (\ref{eq:dt_Te}), (\ref{eq:dt_J}), and (\ref{eq:dt_Pi}) to retrieve

\begin{equation}\label{eq:matrix}
    M X = \bm{0},
\end{equation}
where $\bm{0} = (0,0,0,0,0)^{T}$, $X = (\nabla_{\|} \phi, \nabla_{\|} N_e, \nabla_{\|} T_e, \nabla_{\|} U_{\|}, \nabla_{\|} T_i)^{T}$ and 

\begin{equation} 
M=
    \begin{pmatrix}
        0 & \tau_i T_i + T_e & N_e & N_e U_{\|} & \tau_i N_e \\
        0 & U_{\|} & 0 & N_e & 0 \\
        - N_e & T_e & 1.71  N_e & 0 & 0 \\
        \frac{2}{3}1.71 c_{\phi} N_e T_e & \frac{2}{3}0.71 \left( \upare - U_{\|} \right) T_e & N_e \upare + \frac{2}{3} 1.71 c_{T_e}N_e T_e & -\frac{2}{3}0.71 N_e T_e & 0 \\
        0 & U_{\|} T_i & 0 & \frac{5}{3} N_e T_i & U_{\|} N_e
    \end{pmatrix},\label{eq:Mat_eq}
\end{equation}
with $c_{\phi}=\partial_{\phi}U_{\| e}=- U_{\| e}/T_e$ and $c_{T_e}=\partial_{T_e}U_{\| e}=U_{\| e}\left(1/2+\phi/T_e\right)/T_e$, as it can be deduced directly using \eqref{eq:upare}. 

For a non-trivial solution of \eqref{eq:Mat_eq} to exist, the determinant of $M$ has to vanish. This leads to
\begin{equation} \label{eq:Upari}
    U_{\|} = \pm1.10 \sqrt{T_{e}}\sqrt{1 + 1.37 \frac{\tau_i T_i}{T_e}},
\end{equation}
where the sign is determined by requiring that $U_{\|}$ points towards the wall.

Having determined an inhomogeneous Dirichlet BC for $U_{\|}$, the values the gradients, $\nabla_{\|} \phi$, $\nabla_{\|} N_e$, $\nabla_{\|} T_e$, and $\nabla_{\|} T_i$, are determined by solving the linear problem in \eqref{eq:matrix}. This yields

\begin{subequations}\label{eq:bc_all}
\begin{align}
    \nabla_{\|} N_e &= - \frac{N_e}{U_{\|}} \nabla_{\|} U_{\|}, \label{eq:bcn} \\
    \nabla_{\|} T_i &= - \frac{2T_i}{3U_{\|}} \nabla_{\|} U_{\|}, \label{eq:bcTi} \\
    \nabla_{\|} T_e &= \left(\frac{3T_e + 5 \tau_i T_i}{3 U_{\|}} - U_{\|}\right) \nabla_{\|} U_{\|},\label{eq:bcTe} \\
    \nabla_{\|} \phi &= \left(0.71 \frac{T_e}{U_{\|}} + 1.71 \frac{5\tau_i T_i}{3U_{\|}} - 1.71 U_{\|} \right) \nabla_{\|} U_{\|}. \label{eq:bcphi}
\end{align}
\end{subequations}
The BC for the gyromoments are simply retrieved from the recursion relation in \eqref{eq:Npjrecu}. This implies
\begin{subequations}\label{eq:BCnpj}
    \begin{align}
        \N^{pj}_i&=\begin{cases}
            N_i\frac{\left(-1\right)^j\sqrt{p!}}{\sqrt{2^p}\left(p/2\right)!}\left(T_i-1\right)^{p/2+j},\quad & p \: \text{even}\\
            0, \quad & p \: \text{odd}
        \end{cases}
    \intertext{which, by using the chain rule and Eqs. (\ref{eq:bcn}) and (\ref{eq:bcTi}), yields}
        \nabla_{\|} \N^{pj}_i&=
        -N_i\frac{\left(-1\right)^j\sqrt{p!}}{\sqrt{2^p}\left(p/2\right)!}\left(T_i-1\right)^{p/2+j}\left(1+\frac{T_i}{3\left(T_i-1\right)}\left(p+2j\right)\right)\frac{\nabla_{\|} U_{\|}}{U_{\|}},
\intertext{for even $p$, and}
             \N^{pj}_i&=0,
    \end{align}
\end{subequations}
for odd $p$.

Finally, we use Eqs. (\ref{eq:upare}) and (\ref{eq:Upari}) to obtain the BC for $J_{\|}$ and get
\begin{equation}
    J_{\|}= \pm N_{e}\sqrt{T_{e}} \left[1.1042\sqrt{1 + 1.3669 \frac{\tau_i T_i}{T_e}}-\exp \left(\Lambda - \frac{\phi}{T_e} \right)\right].\label{eq:bcJ}
\end{equation}

We note that, while the structure of our BCs corresponds to that of \cite{Loizu2012} the coefficients in \eqref{eq:Upari} differ slightly (this is due to the inclusion of the $N_e U_{\| e}\nabla_{\|}T_e$ term in \eqref{eq:dt_Te} which was absent in \cite{Loizu2012}). We also note that, by inserting \eqref{eq:Upari} in Eqs. (\ref{eq:bcTe}) and (\ref{eq:bcphi}), the Neumann BCs for $\nabla_{\|}T_e$, and $\nabla_{\|}\phi$ are independent of $T_i$.


For comparison, we also carry out simulations where Bohm fluid boundary conditions are imposed for $U_{\|}$ and $J_{\|}$ \cite{frei2023fullf,mencke2025extended,Loizu2012,Mosetto2015},
\begin{subequations} \label{eq:vparbc}
\begin{align} 
U_{\|}&= \pm \sqrt{T_{e} + \tau_i T_{i} },\\
J_{\|}&=\pm  N_{e}\left(\sqrt{T_{e}+\tau_i T_{i}}-\sqrt{T_{e}} e^{   \Lambda  - \phi / T_{e}}\right), \label{eq:bcJ_hom}
\end{align}
\end{subequations}
while homogeneous Neumann boundary conditions are used all other fields. These simulations have already been presented in previous work, in particular the DK simulations in \cite{mencke_full_delta} consider these BCs. In the following, we refer to the BCs in \eqref{eq:vparbc} as the ad hoc BCs and the BCs in Eqs. (\ref{eq:Upari}) (\ref{eq:bc_all}), (\ref{eq:BCnpj}), and (\ref{eq:bcJ}) as the physical BCs.

\section{Numerical implementation}\label{sec:BC_impl}
To solve the system of Eqs. (\ref{eq:dt_Ne}), (\ref{eq:dt_Te}), (\ref{eq:moment_hierachy_DK}), (\ref{eq:Upar}), (\ref{eq:dt_Omega}), (\ref{eq:Omega_impl}), and (\ref{eq:dt_J}), we use the approach described in Ref. \cite{mencke2025extended,mencke_full_delta} summarized in the following. First, we neglect the $\sim \nabla_\| J_{\|}$ term in \eqref{eq:dt_Te}. Beside being higher order  in $\epsilon_\perp$ in the bulk plasma \cite{mencke_full_delta}, this term gives rise to a nonlinear coupling of the  $\nabla_{\| }T_e$ term in \eqref{eq:dt_J} that is prone to numerical instabilities. We introduce the spatial coordinates, $(x,y,z)$, where $z$ is parallel to the magnetic field of the device, and $x$ and $y$ are perpendicular to $z$ forming a right-handed coordinate system. The $x$ direction is discretized in $N_{x}$ equally-sized grid points and has a total length of $L_x$ such that the grid spacing is $\Delta x=L_x/N_x$. An equivalent discretization is used for $y$ and $z$. All derivatives are calculated with a fourth-order central differences scheme \cite{Giacomin2022}, except for the Poisson bracket operator, $\left[ f,g\right]$, which is evaluated using the fourth-order Arakawa algorithm \cite{Arakawa1997}. A staggered grid approach \cite{Paruta2018} is used in the $z$ direction with $U_{\|}$, $J_{\|}$, and the $\N^{pj}_{i}$ with odd $p$ being evolved on grid points shifted at a distance $\Delta z/2$ with respect to the grid points for $N_{e}$, $T_{e}$, $\phi$, $\Omega$, and $\N^{pj}_{i}$ with even $p$. When advancing the ion model, we evolve a finite number of moments, $\N^{pj}_{i}$, with $p$ from $0$ to $P$ and with $j$ from $0$ to $J$. In \eqref{eq:moment_hierachy_DK}, we set $\N^{pj}_{i}=0$ when $p>P$ and/or $j>J$.

A fifth-order adaptive time-step Runge-Kutta scheme is used to advance all fields \cite{press19922numerical}. Equation (\ref{eq:Omega_impl}) is inverted for $\phi$ using an approach similar to the one used in the two-fluid \verb|GBS| code \cite{Giacomin2022}. For stability reasons, numerical diffusion is added to the right-hand side of all equations for all fields, $f$,

\begin{align}
D(f) = \eta_\perp \left(\partial_{x}^2 +  \partial_{y}^2\right) f + \eta_z \partial_{z}^2 f.
\end{align}

The numerical parameters $\eta_{\perp}>0$ and $\eta_{z}>0$ are chosen as small as possible to guarantee the numerical stability of the simulations, without significantly affecting the simulation results. Finally, for the BCs at the edge of the plasma column, where $\bm B$ is parallel to the simulation boundary, we use $\phi=\Lambda T_e$ and homogeneous Neumann BCs for all other fields. The perpendicular size of the domain is chosen large enough such that these BCs do not significantly affect the plasma dynamics.

We now describe the implementation of the physical BCs in Eqs. (\ref{eq:Upari}) (\ref{eq:bc_all}), (\ref{eq:BCnpj}), and (\ref{eq:bcJ}). We remark that the BCs derived above come in two forms: Dirichlet boundary conditions for $U_{\|}$, and Neumann boundary conditions for all the other quantities. Since derivatives are evaluated using a fourth-order central-differences scheme, two ghost cells are present on each side of the domain in the $z$-direction. We focus on the left wall and note that an equivalent treatment is done for the right wall.

We define the index $l$ for the $z$-grid such that $z=l\Delta z $. We impose that the wall is at $z=0$ and therefore at $l=0$. For $U_{\|}$, we start by imposing that $ U_{\|,0} \equiv - 1.10 c_s\sqrt{1 + 1.37 \frac{\tau_i T_i}{T_e}}$. Then, $U_{\|,-1}$ and $U_{\|,-2}$ are determined by evaluating the forward gradient of $U_{\|}$ at the wall
\begin{subequations}
\begin{align}
    \partial_z U_{\|,0} &\equiv \frac{1}{\Delta z} \left( -\frac{3}{2} U_{\|,0}  + 2 U_{\|,1} - \frac{1}{2} U_{\|,2} \right), \label{eq:gradUpari}\\
\intertext{and extrapolating it linearly into the ghost cells as follows}
    U_{\|,-1} &= U_{\|,0} - \partial_z U_{\|,0} \Delta z, \\
    U_{\|,-2} &= U_{\|,0} - 2 \partial_z U_{\|,0} \Delta z.
\end{align}
\end{subequations}

For all the other fields $f$, the Neumann boundary conditions have the form 
\begin{equation} \label{eq:generalBC}
    \del_z f = g \frac{\del_z U_{\|}}{ U_{\|}},
\end{equation}
where, for example $g=-N_e$ for $f=N_e$, see \eqref{eq:bcn}, and the function $f$ for the other fields are presented in Eqs. (\ref{eq:bc_all}) and (\ref{eq:BCnpj}). To find the values of the two ghost cells, we extrapolate linearly into the ghost domain using the boundary condition in \eqref{eq:generalBC} and the values of $U_{\|,0}$ and $\partial_z U_{\|,0}$ previously computed. More precisely, we define
 \begin{subequations}\label{eq:inhom_neumann}
\begin{align}
    \del_z f_{0} &=  \frac{g_0}{U_{\|,0}} \partial_z U_{\|,0}, 
\intertext{and then impose}
    f_{-1} &= f_0 - \del_z f_0\Delta z, \\
    f_{-2} &= f_0 - 2 \del_z f_0 \Delta z.
\end{align}
 \end{subequations}
We note that the first-order upwind finite-difference scheme for the BCs introduced here makes our simulations robust and stable. Numerical tests show that high-order less-dissipative BCs are less robust. This is the result of large parallel gradients that build up at the sheath entrance and propagate from the sheath to the bulk of the plasma.

\section{Simulation results}\label{sec:sim_res}

We consider a helium plasma in the LAPD linear plasma experiment \cite{Gekelman1991}, with parameters as in Refs. \cite{Rogers2010,frei2023fullf,mencke2025extended}. More precisely, we impose: $n_{e0} = 2 \times 10^{12}\: \mathrm{cm}^{-3}$, $T_{e0} = 6\: \mathrm{eV}$, $T_{i0} = 3 \: \mathrm{eV}$, $\Omega_{i} \sim 960\:\mathrm{kHz}$, $\rho_{s0} = 1.4\:\mathrm{cm}$, $c_{s0}= 1.3 \times 10^{6}\:\mathrm{cm/s}$ , $m_i / m_e =  400$, $\nu_0 = 0.03$, and $\tau_{i}=0.5$. The LAPD vacuum chamber has a radius $R \simeq 0.56\: \mathrm{m}$, therefore we assume $R = 40 \rho_{s0}$ and, with a parallel extension $L_z \simeq 18\: \mathrm{m}$, we impose $L_z=36 R$. With these parameters, the reference time is $R / c_{s0}  \sim 43\:\mu\mathrm{s}$. We set the perpendicular extension of the domain $L_x=L_y=100\rho_{s0}$, sufficiently large for the turbulent structures not to be affected by the perpendicular boundary conditions. For the sources, we choose  $\mathcal{A}_{N\infty} = \mathcal{A}_{T_e \infty}  =  0.004$, $\mathcal{A}_{E\infty} = 0.001$, $\mathcal{A}_{N0}=\mathcal{A}_{Te0}=\mathcal{A}_{E0}  = 0.04$, $L_s = 1  \rho_{s0}$, and $r_s = 20 \rho_{s0}$. 

The number of grid points is $N_x = N_y = 192$, which corresponds to a grid spacing of approximately $ 0.52 \rho_{s0}$ in the perpendicular direction, while $N_z=64$ corresponding to a grid spacing of approximately $ 0.56 R$ in the $z$-direction. For all dynamical fields ($N_e$, $T_e$, $U_{\|}$, $\Omega$, $J_{\|}$, and $\N^{pj}_{i}$) diffusion parameters are chosen as $\eta_z\simeq 4$ and $\eta_{\perp}\simeq 0.5$. Simulations are carried out with $\left(P,J\right)=\left(2,1\right)$, $\left(4,2\right)$, or $\left(6,3\right)$.

We first perform a simulation with $\left(P,J\right)=\left(2,1\right)$ and ad hoc BCs until a quasi-steady state is reached, where the injection of heat and particles from the sources is balanced by the outflow at the sheath boundaries. The simulations with physical BCs derived in \secref{sec:bc} is started using this quasi-steady state as initial conditions and run until a new quasi-steady state is achieved. The values of $\left(P,J\right)$ are then increased.

\begin{figure}
    \centering
    \includegraphics[width=\linewidth]{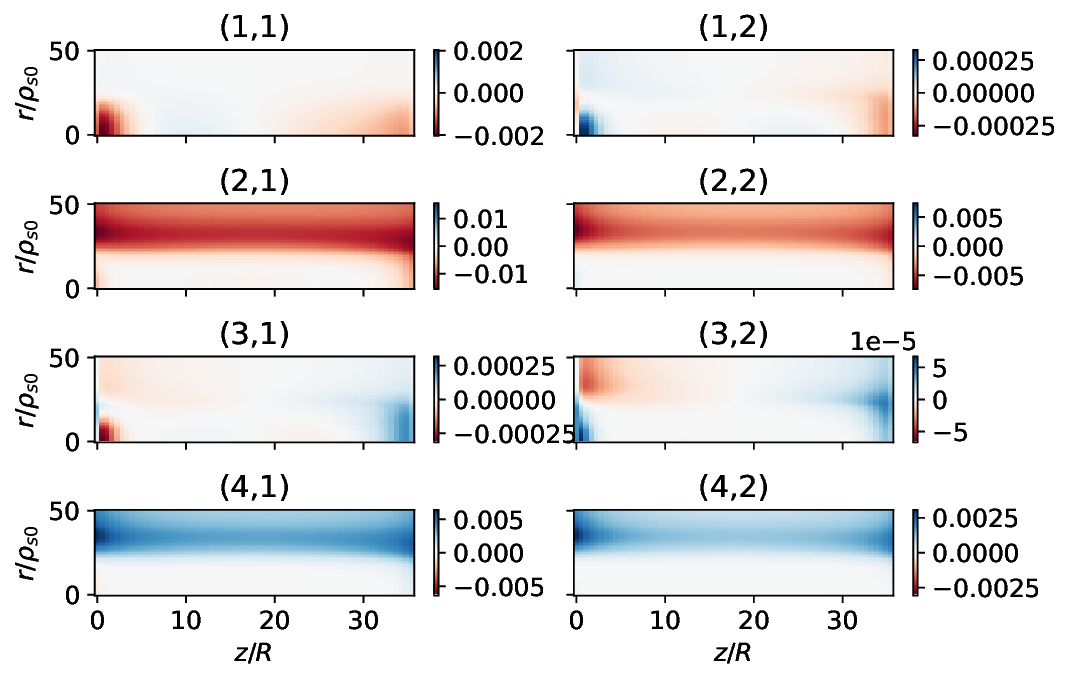}
    \caption{Temporal and azimuthal average of $\N^{pj}_i$ for a simulation with physical BCs and $\left(P,J\right)=\left(6,3\right)$ for different $p=1-4$ (rows) and $j=1,2$ (columns). The moments decrease in amplitude with $p$ and $j$ and the odd-$p$ moments are antisymmetric around $z=L_z/2$, while the even-$p$ moments are symmetric around $z=L_z/2$.}
    \label{fig:npj_avg}
\end{figure}

For presenting the results, we introduce the distance from the axis of the plasma column, $r=\sqrt{x^2+y^2}$ and the azimuthal angle, $\tan \vartheta=y/x$. We define the temporal and azimuthal average of $f$ as follows
\begin{subequations}\label{eq:avg_RMS_SKW}
\begin{align}
    \left\langle f\right\rangle_{t,\vartheta}&=\frac{1}{2\pi}\int_0^{2\pi} d\vartheta\frac{1}{\Delta t}\int_{t_0}^{t_0+\Delta t}d t f,\label{eq:avg_def}
\intertext{the root mean square,}
    RMS\left(f\right)&=\sqrt{\left\langle \left(\left(f\right)-\left\langle f\right\rangle_{t,\vartheta}\right)^2\right\rangle_{t,\vartheta} },
\intertext{and the skewness}
    SKW\left(f\right)&=\left\langle \left(\left(f\right)-\left\langle f\right\rangle_{t,\vartheta}\right)^3\right\rangle_{t,\vartheta}/RMS^3\left(f\right).
\end{align}
\end{subequations}
The temporal and azimuthal average of $\N^{pj}_i$ is shown for different $p$ and $j$ in \figref{fig:npj_avg}. The amplitude of the moments decreases with $j$ and the moments with even $p$ have larger amplitude than those with odd $p$. The rapid decrease of the amplitude of the moments hints that convergence is reached with a small number of moments. We also note that the odd-$p$ moments are antisymmetric around $z=L_z/2$, while the even-$p$ moments are symmetric. Similar observations are reported for the simulations with ad hoc BCs in \cite{mencke_full_delta}. 

\begin{figure*}
    \centering
    \includegraphics[width=\linewidth]{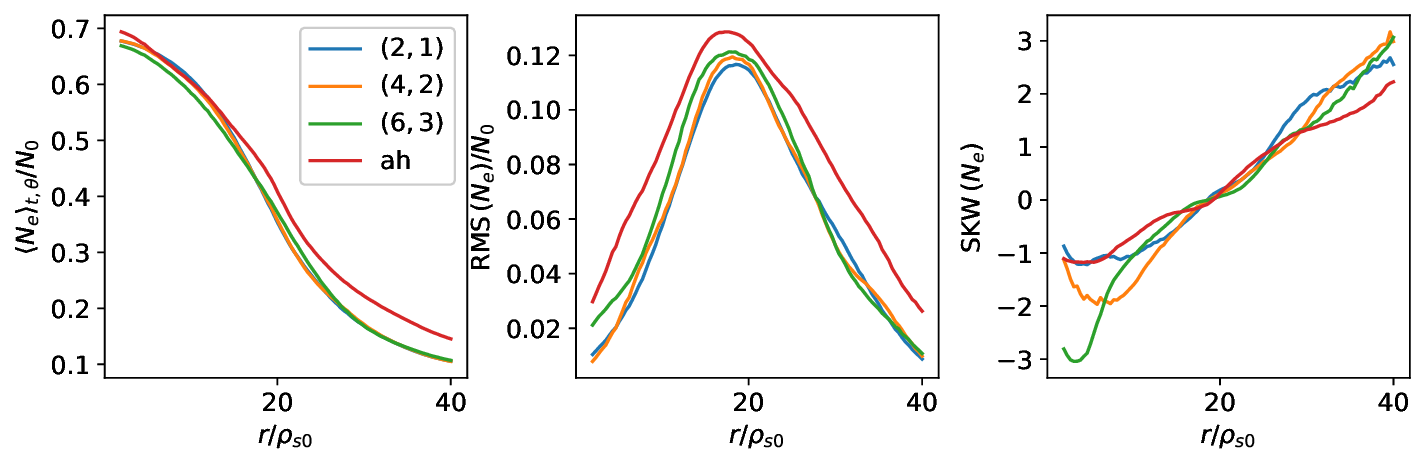}
    \caption{Temporally and azimuthally averaged density, $\left\langle N_e\right\rangle_{\vartheta,t}$ (left), its root mean square, $\operatorname{RMS}\left(N_e\right)$ (center), and skewness  $\operatorname{SKW}\left(N_e\right)$ (right), for simulations with physical BCs and $\left(P,J\right)=\left(2,1\right)$, $\left(4,2\right)$, and $\left(6,3\right)$ and for a simulation with ad hoc BCs with $\left(P,J\right)=\left(6,3\right)$ (ah) at $z=L_z/2$. Fast convergence with $\left(P,J\right)$ is observed. The use of physical BCs leads to a decrease in density for all $r$.}
    \label{fig:ne_conv}
\end{figure*}

\begin{figure}
    \centering
    \includegraphics[width=0.5\linewidth]{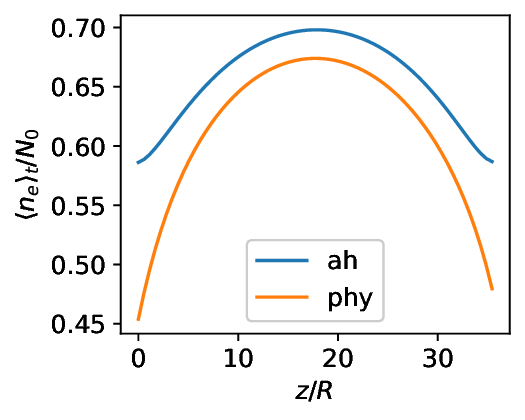}
    \caption{Temporally averaged density, $\left\langle N_e\right\rangle_{t}$ at the centre of the device $\left(x,y\right)=\left(0,0\right)$ for a simulation with $\left(P,J\right)=\left(6,3\right)$ with physical BCs (phy) and ad hoc BCs (ah). A larger decrease of $n_e$ before the sheaths are observed when using physical BCs.}
    \label{fig:par_comp}
\end{figure}

Figure \ref{fig:ne_conv} shows the temporally and azimuthally average of the density as well as its root mean square and skewness for simulations with physical BCs and different choices of $\left(P,J\right)$. A comparison with a simulation using ad hoc BCs is also included. We observe that the simulations with the BCs introduced in \secref{sec:bc} converge with $\left(P,J\right)=\left(2,1\right)$. The convergence of the simulations with ad hoc BCs, as discussed in \cite{mencke_full_delta}, is obtained for $\left(P,J\right)=\left(2,1\right)$ as well. The simulation with the ad hoc BCs has a larger average density and a larger $\operatorname{RMS}$ associated with larger fluctuations. This is due to a reduced outflow resulting from a reduced $U_{\|}$ and a smaller drop in density at the sheath entrance compared to the simulations with the physical BCs. Indeed, \figref{fig:par_comp} confirms that the density decreases more at the sheath entrances when using the physical BCs derived in \secref{sec:bc}. A similar trend is observed for all other fields.

In the quasi-steady state, an average ambipolar outflow of electrons and ions is expected. From \eqref{eq:bcJ}, setting $J_{\|}=0$ at the sheath entrance, we obtain 
\begin{subequations}
\label{eq:quasi_neut}
\begin{align}
    \phi-\Lambda T_{e}&=-T_e\left[\frac{1}{2}\ln \left(1+\frac{T_i}{T_e}\right)+\ln 1.1042\right],\label{eq:quasi_neut_inh}
\intertext{while, when the ad hoc BCs are used, $J_{\|}=0$ yields}
    \phi-\Lambda T_{e}&=-\frac{T_e}{2}\ln \left(1+\frac{T_i}{T_e}\right) ,\label{eq:quasi_neut_hom}
\end{align}
\end{subequations}
by using \eqref{eq:bcJ_hom}.

\begin{figure}
    \centering
    \includegraphics[width=0.5\linewidth]{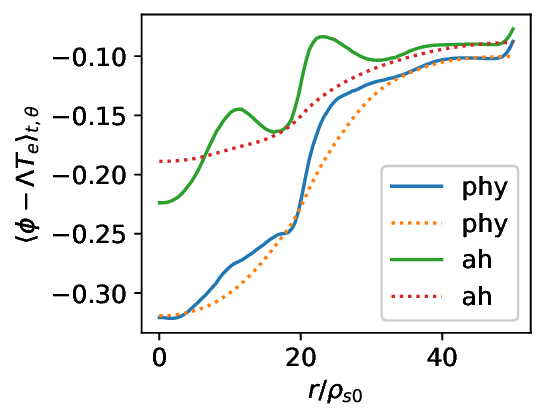}
    \caption{Temporally and azimuthally averaged left-hand sides of \eqref{eq:quasi_neut} (solid) and right-hand sides of \eqref{eq:quasi_neut} (dotted) for a simulation with $\left(P,J\right)=\left(6,3\right)$ and with ad hoc BCs (ah) and physical BCs (phy). In both cases, the ambipolar outflow is well balanced.}
    \label{fig:outflow}
\end{figure}

The left- and right-hand sides of \eqref{eq:quasi_neut} are compared in \figref{fig:outflow}. The relations in \eqref{eq:quasi_neut} are satisfied to a high degree. This reveals that, while the ion outflow for the physical BCs is larger than that of the ad hoc BCs, $\phi/ T_e$ decreases such that the electron outflow matches that of the ions. 

\begin{figure}
    \centering
    \includegraphics[width=0.5\linewidth]{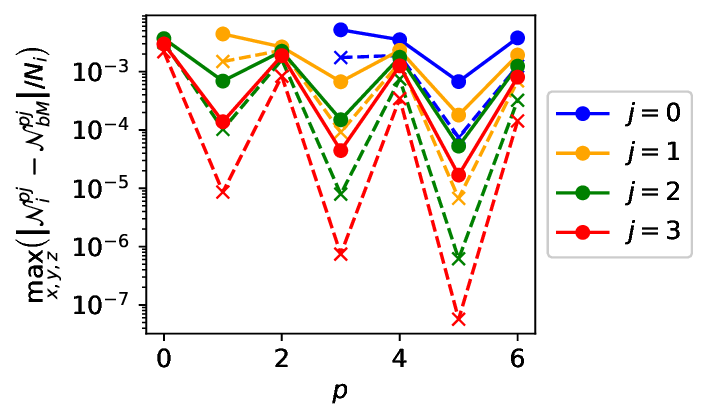}
    \caption{The spatial maximum of the deviation between the gyromoment and those of a shifted biMaxwellian, $\max_{x,y,z}\left(\left|\mathcal{N}_{i}^{pj}-\mathcal{N}_{bM}^{pj}\right|/N_{i}\right)$, for a simulation with $\left(P,J\right)=\left(6,3\right)$ and physical BCs (solid circles) and with ad hoc BCs (dashed crosses). The even-$p$ moments become more biMaxwellian and the odd-$p$ moments increases in magnitude when using inhomogeneous BC instead of homogeneous.}
    \label{fig:Max_deviation}
\end{figure}

Finally, in order to quantify the deviation of the ion distribution function from a local Maxwellian, we compare $\N^{pj}_i$ with those of a local bi-Maxwellian \cite{mencke_full_delta,mencke2025extended}, 
\begin{equation}
        \N^{pj}_{bM}=\begin{cases}
        N_i\frac{\left(-1\right)^{j}\sqrt{p!}}{\sqrt{2^{p}}\left(p/2\right)!}\left(\frac{T_{\| i}}{T_{i0}}-1\right)^{p/2}\left(\frac{T_{\perp i}}{T_{i0}}-1\right)^{j}& p\:\text{even,}\\
        0 & p\:\text{odd,}
    \end{cases}\label{eq:bi_Max}
\end{equation}
in \figref{fig:Max_deviation}. We see that the distribution function is closer to a bi-Maxwellian when using the ad hoc BCs instead of the physical ones. This is a consequence of larger parallel derivatives, when using the physical BCs, as seen from \figref{fig:par_comp}, thus leading to an increase in the terms coupling odd- and even-$p$ moments in \eqref{eq:moment_hierachy_DK}.

Indeed, by studying the steady-state behavior of the moments, we note that parallel derivatives, perpendicular transport associated with the $\bm E\times \bm B$ drift, sources, and the collision operator determine the steady state. Using \eqref{eq:moment_hierachy_DK} setting $\partial_t=0$ the steady state moments, $\N^{pj}_{iss}$, satisfy
\begin{align}
    &\frac{v_{Th i}}{\sqrt{2}}\left(\sqrt{p+1}\nabla_{\|}\N^{p+1j}_{iss}+\sqrt{p}\nabla_{\|}\N^{p-1j}_{iss}\right)+\left(p\N^{pj}_{iss}+\sqrt{p\left(p-1\right)}\N^{p-2j}_{iss}\right) \nabla_{\|}U_{\|}\nonumber\\
    &+\frac{1}{B}\poissonbracket{\phi}{\N^{pj}_{iss}}- \frac{\sqrt{2p}\N^{p-1 j}_{iss}}{N_{ess}}\frac{1}{m_i v_{Thi}}\nabla_{\|} P_{\| i ss}=  C_{ii}^{pj} + S^{pj}_{i}.
\end{align}
We note that only the first and last term on the left hand side act as a source term for the odd $p$ moments since $S^{pj}_{i}=0$ for odd $p$ and the rest of the terms are proportional to an odd $p$ moment. Thus by introducing the inhomogeneous BC and, as a consequence, increasing the parallel derivatives of the even $p$ moments, the odd $p$ moments increase in magnitude.

Finally, we note that the turbulence behavior is not affected by the physical BCs derived in \secref{sec:bc}. Indeed, also with the physical BCs, we observe turbulence with the same properties as in \cite{mencke_full_delta}. The turbulence is driven by a Kelvin-Helmholz instability \cite{Rogers2010,mencke_full_delta}, driven unstable by perpendicular gradients of the electrostatic potential. Thus, a change in the parallel equilibrium does not significantly modify the turbulent drive and, as a consequence, the observed turbulent structures.

\section{Conclusion}\label{sec:concl}
Collisional presheath boundary conditions are derived and implemented for a drift-kinetic gyromoment model in a linear plasma device in the case of perpendicular incidence of the magnetic field to the wall. Imposing that the derivatives in the direction of the sheath dominate, a similar approach as \cite{Loizu2012} is used to derive a set of Dirichlet boundary conditions for the parallel flows and inhomogeneous Neumann boundary conditions for the remaining fields. Extending the fourth order centered finite-difference drift-kinetic simulations in \cite{mencke_full_delta}, we implement these boundary conditions with a robust first order forward finite-difference scheme. 



A larger outflow of particles is observed compared to the ad-hoc homogeneous Neumann boundary conditions used in \cite{mencke_full_delta}. The boundary conditions derived here result in a larger drop in density at the sheath entrance. As a consequence, the odd-$p$ moments increase in amplitude. The outflows of electrons and ions remain ambipolar at the sheath entrances, although the overall outflow increases. Despite the change in the equilibrium profiles, the main turbulence properties are unaffected by the implementation of the physical sheath boundary conditions.




The present work focuses on a singly ionized plasma. Plasma-neutral interactions are neglected. The case of perpendicular incidence is considered. These are all elements that should be generalized in future work. Nevertheless, even in this simplified setting, we show that proper sheath boundary conditions are crucial for determining the equilibrium properties of the plasma.

\section*{Acknowledgements}
The authors thank Alessandro Geraldini, Louis N. Stenger, Brenno Jason Sanzio Peter De Lucca, Samuel Ernst, and Zeno Tecchiolli for fruitful discussions. Simulations were carried out thanks to the grant EHPC-REG-2023R03-144 on the Discoverer supercomputer. This work has been carried out within the framework of the EUROfusion Consortium, partially funded by the European Union via the Euratom Research and Training Programme (Grant Agreement No 101052200 — EUROfusion).
The Swiss contribution to this work has been funded in part by the Swiss State Secretariat for Education, Research and Innovation (SERI).
Views and opinions expressed are however those of the author(s) only and do not necessarily reflect those of the European Union, the European Commission or SERI. 
Neither the European Union nor the European Commission nor SERI can be held responsible for them.

\nocite{*}
\bibliography{biblio}

\end{document}